\documentclass[11pt,a4paper]{article}
\usepackage{graphics}
\usepackage{amsmath}
\usepackage{amsfonts}
\usepackage{amssymb}
\usepackage[margin=1in]{geometry}
\usepackage{epsfig}
\def\eq#1{(\ref{#1})}
\author{Marcelo Ferreira da Silva, \vspace{5mm} Carlos Silva}
\title{The Compton/Schwarzschild duality, black hole entropy and quantum information.}
\begin{document}

%\twocolumn

\maketitle

\begin{abstract}
A new kind of duality has been proposed by Carr related to the quantum description of black holes, the so-called Compton/Schwarzschild duality \cite{Carr:2015nqa}.
In this context, a new form for a Generalized Uncertainty Principle has arisen, which must bring us an interesting new route to a quantum description of
spacetime. In the present paper, we shall investigate the consequences of
the Compton/Schwarzschild duality to black hole entropy. The results found out reinforce an interesting perspective on the relationship between black holes and quantum information theory that has been recently proposed in the literature: that black hole entropy can assume negative values at the final stage of black hole evaporation. Consequently, in the context of the quantum corrections to gravity proposed by the Compton/Schwarschild duality, the final state of a black hole might correspond to a quantum entangled state, in the place of a remnant.

\end{abstract}

\section{Introduction}

The road to a quantum description of the gravitational phenomena has been a tough challenge. In this sense, efforts have been done by different approaches, which have proposed
some interesting results that may underly
possible routes to quantum gravity.
Among such results, we have three interesting hints that have occupied a lot of space in the literature.

The first of such hints is related to dualities. Particularly, dualities consist of an important feature of string theory where we have several examples of them, such as the $T$-duality and the $S$ duality.
Moreover, in string theory context, holographic dualities such as those related to the $AdS/CFT$ correspondence \cite{Maldacena:1997re} have found a large number of applications \cite{Hubeny:2014bla}. Actually, holographic dualities have been shown to be a unifying principle in investigations related to quantum gravity, being a possible bridge between string theory
and Loop Quantum Gravity \cite{Silva:2020bnn, Silva:2020lnr}.

A second interesting hint to a quantum description of the gravitational phenomena is the so-called Generalized Uncertainty Principle (GUP), which is rooted in which must be an elementary feature of quantum gravity: the inclusion of a minimal length scale in the description of the physical world \cite{Konishi:1989wk, Maggiore:1993rv, Maggiore:1993kv, Scardigli:1999jh, Das:2008kaa}. In this way, a GUP replaces the usual Heisenberg Uncertainty Principle, assuming a special role in telling us how the measurement process in quantum gravity must be different from that we have in quantum mechanics.

The third hint to quantum gravity is traced by the Bekenstein-Hawking entropy-area formula which arises in the context of black hole thermodynamics.
Such a formula occupies a central place in the discussions on quantum gravity since it gives us the number of degrees of freedom we have in nature at the most fundamental level. For instance, the Bekenstein-Hawking formula is one of a few equations in physics where the four fundamental constants in nature appear together. Moreover, it establishes a connection between black hole thermodynamics and quantum information theory, giving rise to the concept, by Wheeler, of "it from bit".

%It points to the possibility that the Bekenstein-Hawking formula must be an important tool to obtain a unified physical description of the world.

%In this way, different forms for the GUP can be proposed depending on the quantum gravity approach  we use \cite{}.

Sprouting from such discussions, an interesting result has been introduced by Carr \cite{Carr:2015nqa}, the so-called Compton/Schwarzschild duality. By consisting of a kind of $T$-duality, it says that in three spatial dimensions, the Compton wavelength $(R_C \propto M^{-1})$ and the Schwarzschild radius $(R_S \propto M)$ are dual under the transformation $M \rightarrow M^{2}_{P}/M$, where $M_{P}$ is the Planck
mass. This suggests that there could be a fundamental link between elementary particles
with $M < M_{P}$ and black holes in the $M > M_{P}$ regime.
In the context of such a duality, a new proposal for a GUP has been suggested, which makes it possible to address the two first hints to quantum gravity presented in the paragraphs above through a unique formulation.

In this paper, we shall investigate how one could expand the discussions promoted by the Compton/Schwarzschild duality to include the third hint we have pointed out above, i.e., the Bekenstein-Hawking entropy/area formula. In this way, we shall study black hole thermodynamics in the context of the Compton/Schwarzschild duality, by considering quantum corrections coming from the GUP that arise in the context of such a duality. In this way, we obtain a quantum corrected formula to black hole entropy which possesses a distinct behavior when compared with the usual Bekenstein-Hawking formula. The most intriguing feature of the result obtained in the present paper will be the possibility of having negative values for the black hole entropy at the final stages of the black hole evaporation when the black hole enters the subplanckian regime. In this case, the final state of a black hole could correspond to an entangled quantum state where the quantum information related to the black hole's initial state could be stored. It corresponds to a different scenario that has been proposed by other GUP approaches, where the final stage of a black hole corresponds to a remnant \cite{Adler:1999bu, Adler:2001vs, Chen:2002tu, Chen:2003bn}.

The paper is organized as follows: in section \eq{comp/schw}, we shall review the Compton/Schwarzschild duality as it has been introduced by Carr \cite{Carr:2015nqa}. In section \eq{quantum-area-law}, we shall derive a quantum corrected Bekenstein-Hawking formula in the context of the Compton Schwarzschild duality. Section \eq{conclusions} is devoted to conclusions and perspectives.

\section{The Compton/Schwarzschild duality and the GUP}\label{comp/schw}

By considering a microscopic description of reality, quantum mechanics says that a central role is assumed by the Heisenberg Uncertainty Principle, which gives us that the uncertainties in the position $\delta x$ and in the momentum $\delta p$ of a particle must obey the relation $\delta x \geq \hslash / 2\delta p$. It implies that, if the momentum of a particle has an upper bound given by $mc$, then it is not possible to localize the particle in a region smaller than $\hslash /2mc$. It defines a characteristic length related to the particle, the so-called reduced Compton length, given by $\hslash /mc$.

On the other hand, by considering a macroscopic description of the world, black holes assume a key role. Black holes consist of a region of spacetime where the gravitational field is so strong that not even light can escape from there. The characteristic length of such a region is the black hole Schwarzschild radius $R_{S} = 2Gm/c^{2}$, where $m$ is the black hole mass, which defines the black hole boundary, called the event horizon. Such a putative solution of General Relativity has been considered a gateway for quantum gravity, and it has gotten more attention in the last years due to the detection of gravitational waves by LIGO \cite{LIGOScientific:2016aoc}, and the capturing of the first black hole image by the Event Horizon Telescope \cite{EventHorizonTelescope:2019dse}.

It has been suggested that the Schwarzschild radius and the Compton length may intercept around the Planck scale. However, the way it can occur is not well understood yet, since both quantum mechanics and general relativity must break in such a regime.
In the sense of shedding more light on this issue, it was proposed by Adler and colleagues that at the Planck scale the Heisenberg Uncertainty Principle must be replaced by a GUP written as \cite{Adler:1999bu, Adler:2001vs, Chen:2002tu, Chen:2003bn}

\begin{equation}
\delta x \geq \frac{\hslash}{\delta p} + \alpha l_{P}^{2}\Big(\frac{\delta p}{\hslash}\Big)\;, \label{adler-gup}
\end{equation}

\noindent where $\alpha$ is an adimensional constant (normally considered positive), and $l_{P}$ is the Planck length. In the Eq. \eq{adler-gup} above, the first term on the right-hand side is given by the quantum mechanical contribution, while the second term corresponds to the gravitational contribution to the GUP.

The GUP in the Eq. \eq{adler-gup}, where the quantum mechanical and gravitational contributions are added linearly,
bring us a crucial consequence for black hole thermodynamics, especially for the final stage of black hole evaporation, where quantum gravity effects must become more evident.
For example, in the context of such a GUP, a black hole is prevented to evaporate completely, in a similar way the hydrogen atom is avoided from collapsing by the usual Heisenberg uncertainty principle. In this way, a black hole remnant must be found at the end of the evaporation process. \cite{Adler:2001vs, Chen:2002tu, Chen:2003bn}. Such results have launched several interesting ideas, e.g., the one that black holes could correspond to the origin of dark matter \cite{ Chen:2002tu, Chen:2003bn}.

The possibility of a connection between the Schwarzschild radius and the Compton length at the Planck scale has also motivated the proposition of an interesting duality by Carr \cite{Carr:2015nqa}, the so-called Compton/Schwarzschild duality. By consisting of a kind of $T$-duality, it says that in three spatial dimensions, the Compton wavelength $(R_C \propto M^{-1})$ and Schwarzschild radius $(R_S \propto M)$ are dual under the transformation $M \rightarrow M^{2}_{P}/M$, where $M_{P}$ is the Planck
mass. This suggests that there could be a fundamental link between elementary particles
with $M < M_{P}$ and black holes in the $M > M_{P}$ regime.

However, even though it is possible to find some motivation for such a duality from the GUP \eq{adler-gup}, it has been argued by Carr that, since the contributions from the Compton length and the Schwarzschild radius are independent, it is more natural to assume that they must be added quadratically:

\begin{equation}
\delta x \geq \sqrt{\Big(\frac{\hslash}{\delta p} \Big)^{2} + \Big(\frac{\alpha l_{P}^{2}\delta p}{\hslash} \Big)^{2}} \; . \label{compton-shwarzchild-gup}
\end{equation} 

\vspace{5mm}

It introduces a completely different perspective about the role of quantum corrections coming from a GUP to the gravitational phenomena.
In the next section, we shall address how such a version of a GUP can be connected to a new insight that has been launched in the literature: that black holes can store negative entropy \cite{Song:2014}.
It will be shown that differently than we have in the context of the GUP introduced by Adler et al, the final stage of a black hole might not correspond to a remnant, but an entangled quantum state with a negative entropy associated with it.

%\section{A GUP from the Compton/Schwarzschild duality} 

\section{Quantum corrected Bekenstein-Hawking formula from the Compton/Schwarzschild duality}\label{quantum-area-law}

Let us consider the Schwarzschild black hole, whose metric is

\begin{equation}
ds^2 = -(1 - \frac{2M}{r})dt^2 + (1 - \frac{2M}{r})^{-1} dr^2 + r^2 d\Omega^2_2.    
\end{equation}

For a black hole that absorbs and emits particles of energy $dM \approx c \delta p$, the increase (decrease) in the horizon area can be expressed through

\begin{equation} \label{area}
dA = 8 \pi r_h dr_h = 32 \pi M dM.
\end{equation} 

Moreover, the black hole radiation is a quantum effect. Thus it must satisfy the Heisenberg uncertainty relation:

\begin{equation} \label{heisenberg}
\delta p_i \delta x_j \geq \delta_{ij}.
\end{equation}

However, in the cases where gravity becomes important, the Heisenberg principle should be replaced by a GUP. More specifically, here we shall consider the GUP proposed by Carr: 

\begin{equation} \label{gup}
\delta x^2 \geq \left( \frac{\hbar}{\delta p} \right)^2 + \left( \alpha l_p^2 \delta p / \hbar \right)^2. 
\end{equation}

\vspace{5mm} 

\noindent From the expression above, we get the momentum uncertainty for the emitted particle:

\begin{eqnarray} 
\delta p = \sqrt{\frac{\delta x^2 \hbar^2 \pm \hbar^2 \sqrt{\delta x^4 - 4 l_p^4 \alpha^2}}{2 l_p^4 \alpha^2 }} \\
= \pm \frac{\hbar}{2 \alpha l_p^2} \delta x \sqrt{2 - 2 \sqrt{1 - \frac{4 \alpha^2 l_p^4}{\delta x^4}}}, \label{deltap}
\end{eqnarray}

\noindent where we have taken the minus signal inside the square root to recover the Heisenberg principle at the semi-classical limit.

The expansion of the Eq. \eqref{deltap} gives us

\begin{eqnarray}
\delta p = \frac{\hbar \delta x}{2 \alpha l_p^2} \sqrt{2 - 2 \left( 1 - \frac{2 \alpha^2 l_p^4}{(\delta x)^4} - \frac{2 \alpha^4 l_p^8}{(\delta x)^8} - \dots \right)} \\
= \frac{\hbar}{\delta x} \sqrt{\left[ 1 + \frac{\alpha^2 l_p^4}{(\delta x)^4} + \dots \right]} \;\;.
\end{eqnarray}

\vspace{5mm} 

We have, from \ref{area} and \ref{heisenberg}, that the change in the area of the black hole horizon can be written as (for $c=1$):

\begin{equation} \label{da2}
dA = 32 \pi M dp = 32 \pi M \frac{1}{\delta x},
\end{equation} 

\noindent which is related to a change of black hole entropy given by

\begin{equation} \label{da2}
dS = 8 \pi M dp = 8 \pi M \frac{1}{\delta x},
\end{equation} 

On the other hand,  the results coming from the GUP proposed by Carr give us the quantum corrected expression for the change of the black hole entropy as

\begin{equation}
d S_g = 8 \pi M dp = 8 \pi M \frac{1}{\delta x} \sqrt{\left[ 1 + \frac{\alpha^2 l_p^4}{(\delta x)^4} + \dots \right]} = \sqrt{\left[ 1 + \frac{\alpha^2 l_p^4}{(\delta x)^4} + \dots \right]} dS ,
\end{equation}

\noindent where we have used $\hbar = 1$.

Now, by taking $\delta x \approx 2 r_s = 2 \sqrt{A/4\pi} = \sqrt{S/\pi}$, we shall have
 
\begin{equation} \label{dA}
d S_g = \sqrt{\left( 1 + \frac{16\pi^2 \alpha^2 l_p^4}{S^2} + \dots \right)} dS \;.
\end{equation}

By neglecting higher-order terms, we can integrate \eqref{dA} to find

%\begin{equation} \label{Ainteg}
%A_{gup} = \sqrt{A^2 + \pi^2 \alpha^2 l_p^4} + \frac{\pi \alpha l_p^2}{2} \left( \ln{\left[\frac{\sqrt{A^2 + \pi^2 \alpha^2 l_p^4}}{\pi \alpha l_p^2}  -1 \right]}  - \ln{\left[\frac{\sqrt{A^2 + \pi^2 \alpha^2 l_p^4}}{\pi \alpha l_p^2} + 1  \right]} \right) .
%\end{equation}  

\begin{equation} \label{2entrop}
S_{gup} = \frac{\sqrt{A^2 + \pi^2 \alpha^2 l_p^4}}{4} + \frac{\pi \alpha l_p^2}{8} \left( \ln{\left[\frac{\sqrt{A^2 + \pi^2 \alpha^2 l_p^4} }{\pi \alpha l_p^2} -1  \right]}  - \ln{\left[\frac{\sqrt{A^2 + \pi^2 \alpha^2 l_p^4}}{\pi \alpha l_p^2}  + 1 \right]} \right). 
\end{equation}

The equation above corresponds to the quantum corrected Bekenstein-Hawking entropy-area law in the context of the Compton/Schwarzschild duality.
As one can notice, such an expression will correspond to the classical law when one takes $\alpha = 0$. 

Below, we plot the entropy \eqref{2entrop} as a function of the black hole mass, with three different values of $\alpha$, and considering $l_p = M_{p} = 1$, where $M_{p}$ is the Planck mass. The red line corresponding to $\alpha = 0.5$, the blue line corresponding to $\alpha = 0.25$ and the yellow line corresponding to $\alpha = 0$.

%\begin{tikzpicture}
%
%\begin{axis}[
%xlabel = $S(A)$,
%ylabel = {$S_{gup}(S)$},
%]
%\addplot[blue,domain = 0:10] {sqrt(x^2 + 2.46740) + 0.78539*(- ln(sqrt(2.46740 + x^2)/1.57079 + 1) + ln(sqrt(2.46740 + x^2)/1.57079 - 1))};
%\addlegendentry{\(\alpha = 0.5\)}
%
%\addplot[red,domain = 0:10] {x};
%\addlegendentry{\(\alpha = 0\)}
%
%\addplot[green,domain = 0:10] {sqrt(x^2 + 9.86960) + 1.57*(- ln(sqrt(9.8696 + x^2)/3.1415 + 1) + ln(sqrt(9.8696 + x^2)/3.1415 - 1))};
%\addlegendentry{\(\alpha = 1\)}
%
%\end{axis}
%\end{tikzpicture}

%%%%%%%%%%%%%%%%%%%%%%%%%%%%%%%%%%%%%%%%%%%%%
%%%%%%%%%%%%%%%%%%%%%

\begin{figure}[htb]
     \centering  % figura centralizada
      \includegraphics[width=10cm, height=7cm]{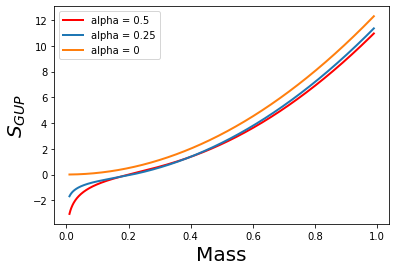}
      \caption{Black hole entropy in terms of the BH mass for the sub-Planckian regime. } \label{fig1}
\end{figure}

%\begin{tikzpicture}
%
%\begin{axis}[
%xlabel = $M$,
%ylabel = {$S_{gup}(S)$},
%]
%\addplot[blue,domain = 0:2] {0.39270*sqrt(16*(pi^2)*x^4 + 2.46740) + 0.19635*(- ln(sqrt(2.46740 + 16*(pi^2)*x^4)/1.570796 + 1) + ln(sqrt(2.46740 + 16*(pi^2)*x^4)/1.570796 - 1))};
%%\addlegendentry{\(\alpha = 0.5\)}
%
%\addplot[red,domain = 0:2] {4*pi*x^2};
%%\addlegendentry{\(\alpha = 0\)}
%
%\addplot[green,domain = 0:2] {0.78540*sqrt(16*(pi^2)*x^4 + 9.86960) + 0.39270*(- ln(sqrt(9.8696 + 16*(pi^2)*x^4)/3.1415 + 1) + ln(sqrt(9.8696 +16*(pi^2)*x^4)/3.1415 - 1))};
%
%
%%\addplot[brown,domain = 0:2] {0.78540*sqrt(16*(pi^2)*x^4 + 9.86960) - 0.39270*(\arctanh(sqrt(9.8696 + 16*(pi^2)*x^4)/3.1415)};
%
%
%
%\addplot[brown, dashed, domain = 0:2] {0};
%%\addlegendentry{\(\alpha = 1\)}
%
%%\addplot[orange,domain = 0:2] {4*pi*((x^2)*(2- sqrt(1-x^-2)) - ln(x + sqrt(x^2 -1)))};
%
%%\addlegendentry{\(\alpha = 1\)}
%
%\end{axis}
%\end{tikzpicture}

%%%%%%%%%%%%%%%%%%%%%%%%%%%%%%%%%%%%%%%%%%%%%
%%%%%%%%%%%%%%%%%%%%%%%%%

From Fig. \eq{fig1}, we see that in the sub-Planckian regime, we have a different behavior of the black hole entropy as compared with the classical entropy, with a bigger deviation for bigger values of $\alpha$.
It is very interesting that, in the subplanckian regime, negative values for the BH entropy have been obtained.

\section{Conclusions and Perspectives}\label{conclusions}

%In the context of quantum gravity investigations, holographic dualities, GUP  have an important hole. In such a context, a new GUP has been obtained by Car from the so-called Compton/Schwarzchild duality \cite{Carr:2015nqa}.

What is the final stage of a black hole? If one considers the usual Bekenstein-Hawking scenario for black hole evaporation, the final stage of a black hole must correspond to the that it completely evaporates into thermal radiation.
On the other hand, in the scenario proposed by Adler et al, where quantum gravity corrections due to a GUP are taken into account, the final stage of a black hole must correspond to a remnant.

In the present paper, we have obtained a different scenario. By considering the Compton/Schwarzschild duality proposed by Carr \cite{Carr:2015nqa}, we have obtained an intriguing result where black holes can possess negative entropies at the final stages of their evaporation.
Such a feature of black hole entropy found in the present paper appears as a surprising and counter-intuitive feature of the quantum description of the black hole evaporation process that
is unparalleled in classical information theory. On the other hand, the concept of negative entropy has assumed an important role in quantum information theory, where a negative entropy is related to a quantum entangled state \cite{Song:2014, Cerf:1995sa, Rio:2011}.

In this way, the results found in the present paper might point to the fact that, at the final stages of black hole evaporation, the black hole must enter into a pre-spacetime regime where the information related to the initial state of the black hole must be stored not in a remnant, as occurs in the Adler approach, but into a  quantum entangled state.

{ 
 }
\end{document}